\documentclass[a4paper,11pt]{article}
\newcommand*{\cventry}[7][.25em]{
  \noindent\begin{tabular*}{\textwidth}{l@{\extracolsep{\fill}}r}%
	  {\bfseries #4} & {\bfseries #5} \\%
	  {\itshape #3\ifthenelse{\equal{#6}{}}{}{, #6}} & {\itshape #2}\\%
  \end{tabular*}%
  \ifx&#7&%
    \else{\\\vbox{\small#7}}\fi%
  \par\addvspace{#1}}

\newcommand*{\hintfont}{\bfseries}
\newcommand*{\hintstyle}[1]{{\noindent\hintfont{#1}}}
\newcommand*{\cvitem}[3][.25em]{%
  \ifthenelse{\equal{#2}{}}{}{\hintstyle{#2}: }{#3}%
  \par\addvspace{#1}}

\let\OriginalQuotation\quotation
\renewcommand*{\quotation}{\OriginalQuotation\small\sf}

\newcommand{\comment}[1]{}
\newcommand{\be}{\begin{equation}}
\newcommand{\ee}{\end{equation}}
\newcommand{\ba}{\begin{array}}
\newcommand{\ea}{\end{array}}
\newcommand{\baa}{\begin{array}}
\newcommand{\eaa}{\end{array}}
\newcommand{\bea}{\begin{eqnarray}}
\newcommand{\eea}{\end{eqnarray}}

\newcommand{\MS}{{\overline{\rm MS}}}
\newcommand{\tl}{\tilde l}
\newcommand{\tL}{\tilde L}

\newcommand{\mupt}{\mu_{\rm pt}}

\newcommand{\lambdat}{ \lambda_{\rm TGF}}

\newcommand{\LTGF}{ \Lambda_{\rm TGF}}

\usepackage{array}
\newcolumntype{\Vert}{!{\vrule width 1pt}}

\newcommand{\ssfill}{\xleaders\hbox to 0.35em{\scriptsize.}\hfill}
\newcommand{\oset}[3][0ex]{%
	\mathrel{\mathop{#3}\limits^{
			\vbox to#1{\kern-2\ex@
				\hbox{$\scriptstyle#2$}\vss}}}}

\usepackage{pos}
\usepackage{subcaption}
\title{The SU(N) running coupling in the twisted gradient flow scheme and volume independence}
\ShortTitle{The SU(N) running coupling in the twisted gradient flow scheme}

\author*[ab]{Jorge~Dasilva~Gol\'an}
\author[a]{Margarita~Garc\'{\i}a~P\'erez}
\author[c]{Alberto~Ramos}

\affiliation[a]{Instituto de F\'{\i}sica Te\'orica UAM-CSIC, Nicol\' as Cabrera 13-15,
    Campus de Cantoblanco.\\
    28049, Madrid, Spain}

\affiliation[b]{Departamento de F\'{\i}sica Te\'orica, m\'odulo 15, Universidad Aut\'onoma de Madrid, Cantoblanco.\\
    28049, Madrid, Spain}

\affiliation[c]{Instituto de F\'{\i}sica Corpuscular (IFIC), CSIC-Universitat
    de Valencia.\\
    46071, Valencia, Spain}

\emailAdd{jorge.dasilva@uam.es}
\emailAdd{margarita.garcia@uam.es}
\emailAdd{alberto.ramos@ific.uv.es}

\abstract{We report on an ongoing study of the running coupling of SU(N) pure Yang-Mills
theory in the twisted gradient flow scheme (TGF). The study exploits the idea 
that twisted boundary conditions reduce finite volume effects, leading to an 
effective size in the twisted plane that combines the number of colours and the 
torus period. We test this hypothesis by computing the TGF running coupling and 
the SU(N) $\Lambda-$ parameter on asymmetric lattices of size $(NL)^2L^2$ for 
various gauge groups. Finite volume effects are monitored by analyzing the 
coupling in different planes and by comparing results at different number of 
colours.}

\FullConference{%
 The 38th International Symposium on Lattice Field Theory, LATTICE2021
  26th-30th July, 2021
  Zoom/Gather@Massachusetts Institute of Technology
}

\begin{document}
\begin{flushright}
  IFT-UAM/CSIC-21-139  \\
  IFIC/21-48
\end{flushright}

\maketitle

\section{Introduction}
\label{sec:introduction}
The strong coupling $\alpha_s$ is a key quantity in high energy physics, and throughout the years its determination up to high level of precision has been an important task with theoretical and phenomenological implications. In a recent work, it has been shown that a precise determination of $\alpha_s$ can be made via a non-perturbative matching between QCD and the pure gauge theory using heavy quarks~\cite{DallaBrida:2019mqg}. This has revived the interest in the calculation of the $\Lambda-$parameter in the pure gauge theory. The twisted gradient flow scheme (TGF) introduced in~\cite{Bribian:2019ybc,Bribian:2021cmg,Ramos:2014kla} has several advantages that make it well suited for this purpose, among them a reduced memory footprint that allows for a better usage of GPU clusters. In addition to high precision determinations in the case of $SU(3)$, this scheme is especially suited for the extraction of the $N$ dependence of the $\Lambda-$parameter. In this work we will summarise our recent results on $SU(3)$~\cite{Bribian:2021cmg} and present our ongoing work on the study of the dependence with the number of colours. 

This renormalization scheme is based in 3 main ingredients. The first one is a coupling definition based in the gradient flow (GF)~\cite{Narayanan:2006rf,Lohmayer:2011si,Luscher:2009eq}, computable on the lattice with high precision. Gauge invariant composite observables based on the GF are automatically renormalized quantities for $t>0$ at every order in perturbation theory~\cite{Luscher:2011bx}, so its effect is basically removing ultraviolet divergences and smoothing the gauge field in a range scale $\sqrt{8t}$, that becomes a natural renormalization scale. 

The other two ingredients are referring to the geometry. The TGF is defined by introducing $SU(N)$ Yang-Mills theories on an asymmetric torus of size $(Nl)^2\times l^2$. The short directions are endowed with twisted boundary conditions~\cite{tHooft:1979rtg} satisfying
\begin{equation}
	A_\mu (x + l \hat \nu ) = \Gamma_\nu A_\mu (x) \Gamma_\nu^\dagger , \text{ for } \nu  = \text{1 or 2;}\hspace{0.5cm}(\Gamma_1 \Gamma_2 = Z_{12} \Gamma_2 \Gamma_1)
\end{equation}
where $\Gamma_\mu$ are constant $SU(N)$ matrices and $Z_{12}=\exp{(2\pi i k/N)}$ (for k, $N$ coprime integers) is an element of the $SU(N)$ center. In the long directions gauge fields are periodic with period $\tl\equiv N l$. 

Those ingredients (coupling based on the GF, twisted boundary conditions and the asymmetric geometry) lead us to the following coupling definition (for a precise definition of the normalization factor ${\cal A}(\pi c^2)$, see~\cite{Bribian:2021cmg}):
\begin{equation}
        \lambdat(\mu) =\frac{ 128 \pi^ 2 t^2 }{3 N  {\cal A}(\pi c^2)} \frac{\langle E\left(t\right)\delta_Q\rangle}{\langle \delta_Q\rangle} \Big |_{_{\sqrt{8t}=c \tl = \mu^{-1}}}.
        \label{eq:lambdat}
\end{equation}

Notice that in the context of finite size scaling, the renormalization scale $\mu=1/\sqrt{8t}$ is related with the finite size of the box by taking $\sqrt{8t}=c\tilde{l}$, with $c$ a real, smaller than 1, parameter characterizing the scheme. Perturbation theory indicates that in the large $N$ limit the finite volume effects are controlled by $\tilde{l}$~\cite{Perez:2018afi,Perez:2013dra,Bribian:2019ybc}, rather than $l$, so it is natural to use $\tilde{l}$ to set the scale of the running coupling. In this work, we have set $c=0.30$. The $\delta_Q$ function in equation~\eqref{eq:lambdat} projects the path integral into the sector of configurations with zero topological charge.  This projection aims to circumventing the problem of topology freezing on the lattice~\cite{Fritzsch:2013yxa}.

\section{The $\Lambda-$parameter}
\label{sec:lambda-parameter}
\label{lambda}
One of the aims of this work is to test the validity of the TGF scheme as a tool for high precision calculations, and this has been done by computing the $\Lambda-$parameter of the pure $SU(3)$ YM theory. This parameter is a renormalization group invariant and does not depend on the scale, but it depends on the renormalization scheme. The coefficient that relates $\Lambda$ in the TGF scheme with the $\MS$ scheme can be determined in perturbation theory and has the following value
\be
\log\left(\frac{\LTGF}{\Lambda_\MS}\right)=\frac{3}{22}\left(\frac{11}{3}\gamma_E+\frac{52}{9}-3\log3+C_1(c)\right),
\label{eq:LTGF_LMS} 
\ee where $C_1(c)$ has been computed in~\cite{Bribian:2019ybc} for several values of $c$ and different gauge groups. For the particular cases to be discussed in this work, $C_1(c=0.3)=0.508(4)$ for $SU(3)$ and twist $k=1$, and $C_1(c=0.3)=0.597(14)$ for $SU(5)$ and twist $k=2$. 

Starting from the integration of the RG equations in some renormalisation scheme s
\be
\beta_{\rm s}(\lambda_{\rm s}) = \frac{{\rm d} \lambda_{\rm s}}{{\rm d} \log(\mu^2)},
\label{eq:betaf} 
\ee $\Lambda$ can be determined by taking the limit:
\begin{equation}
  \frac{\Lambda_{\rm s}}{\mu_{\text{ref}}} = \lim_{\lambda_s(\mupt) \to 0}
  \left(b_0 \lambda_s(\mupt)\right)^\frac{-b_1}{2{b_0}^2}
\exp\left(\frac{-1}{2b_0 \lambda_s(\mupt)} \right) I_s^{(n)}\left(\lambda_s(\mupt)\right)
  \times
\exp\left[-\int_{\lambda_s(\mupt)}^{\lambda_s(\mu_{\text{ref}})} \frac{dx}{2 \beta_{\rm s}(x)}
\right],
\label{eq:lambdalim}
\end{equation}
where $\mu_{\text{ref}}$ is a hadronic reference scale, $\mupt$ is a high energy scale where perturbation theory can be applied, and $I_s^{(n)}$ has the following form:
\be I_{\rm s}^{(n)}(\lambda) = \exp \left
\{-\int_0^{\lambda} dx \left(\frac{1}{2 \beta_{\rm s}^{(n)}(x)}+\frac{1}{2 b_0
x^2}-\frac{b_1}{2 b_0^2x}\right)\right \},
\label{eq:Ioflambda} 
\ee
with $\beta_{\rm s}^{(n)}(x)$ the $n-$loop $\beta$-function.

The last factor of equation~\eqref{eq:lambdalim} can be determined non-perturbatively in the lattice by simulating the running of the coupling constant $\lambda_{\rm s}$. We can compute this factor using the step scaling function
\begin{equation}
  \label{eq:sigmadef}
  \sigma(u) = \lambda_s(\mu/2)\Big|_{\lambda_s(\mu) = u},
\end{equation}
that changes the renormalisation scale by a factor of 2 each application and runs the coupling as $u_{k} =\lambda_s(2^k\mu_{\text{ref}})$ up to perturbative scales where perturbation theory can be applied at the matching point $\lambda_{\text{PT}}$. Details of the lattice determination of this quantity in the TGF scheme can be found in~\cite{Bribian:2021cmg}. One of the steps involves taking the continuum limit of the lattice determined step scaling function $\Sigma(u, \tilde L)$, where $\tilde{L}=\tilde{l}/a$. Several alternatives can be taken for this, as for instance fitting raw data to a global fit of the form
\begin{equation}
  \label{eq:globalfit}
\frac{1}{\Sigma(u, \tilde L)} = \frac{1}{u} -2b_0 \log 2 - 2b_1 u \log 2 +
\sum_{k=2}^{4} p_k u^k +
\left(  \sum_{k = 0}^{4} \rho_k u^k \right) \times \frac{1}{\tilde L^2},
\end{equation}
where $\tilde{L}=N\times L$ and $\Sigma(u, \tilde L)$ is determined by measuring the coupling at size $\tilde{L}$ and $2\tilde{L}$ at the same value of bare coupling. Then, the continuum step scaling function $\sigma(u)$ can be computed by taking the $\tilde{L}\rightarrow\infty$ limit of $\Sigma(u, \tilde L)$.

\begin{figure}[t] \centering
\includegraphics[width=0.8\textwidth]{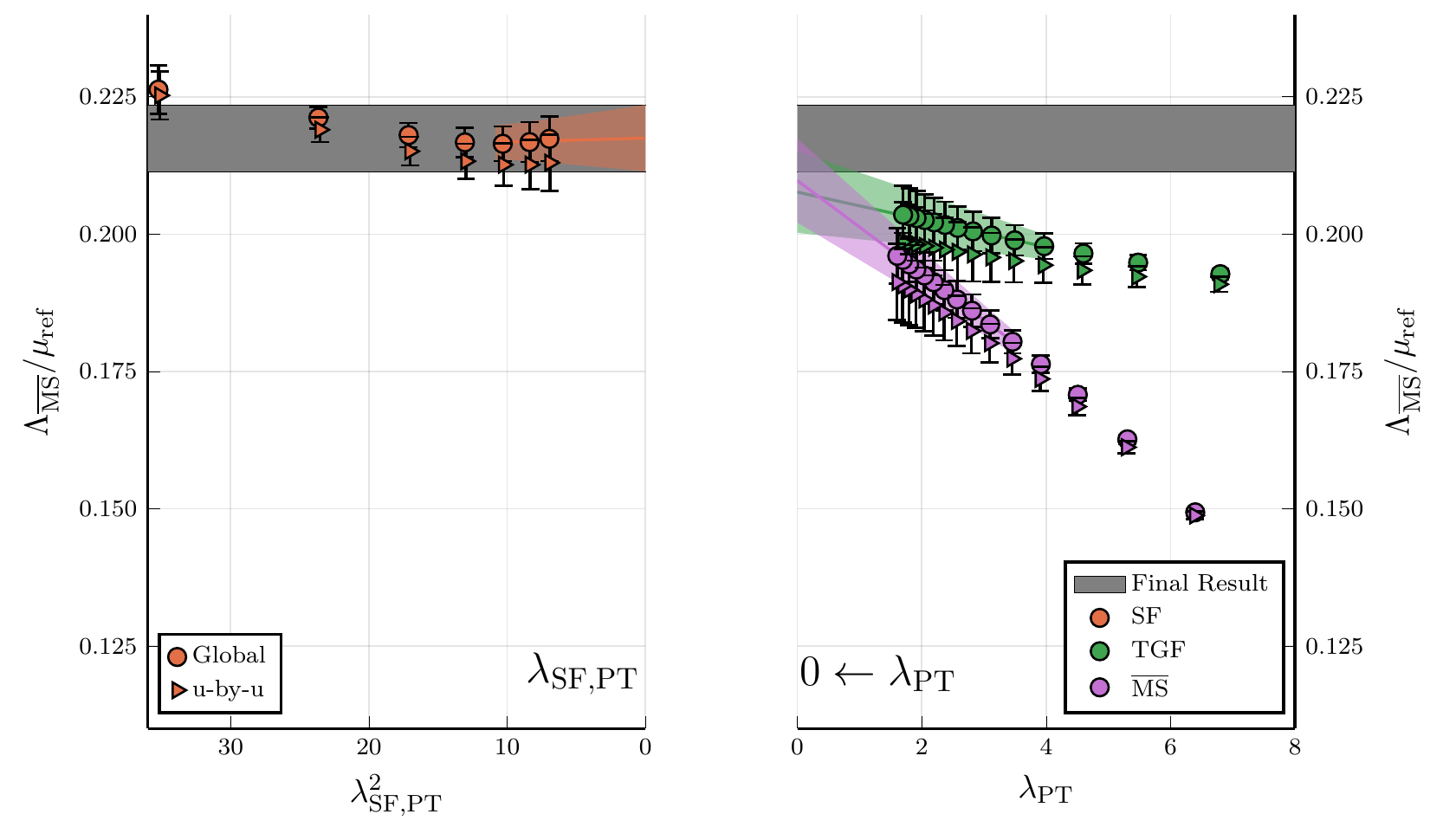}
\caption{$\Lambda_\MS/\mu_{\text{ref}}$ as a function of the matching scale $\lambda_{\text{PT}}$ with perturbation theory, computed throughout several methods, including TGF, perturbative matching with the $\MS$ scheme and non-perturbative matching with the SF scheme.}
\label{comparative}
\end{figure}

An illustration of the validity of this method is presented in fig.~\ref{comparative} for the case of $SU(3)$, where we are comparing the extrapolation of $\Lambda_\MS/\mu_{\text{ref}}$ using the TGF method explained above with the results obtained by the non-perturbative matching to the Schrodinger functional data of~\cite{DallaBrida:2019wur} or through perturbative matching to the $\MS$ scheme, see~\cite{Bribian:2021cmg} for details. We have fixed $\mu_{\text{ref}}$ through the renormalisation condition $\lambda(\mu_{\text{ref}})\equiv 13.9164955$.

\begin{figure}[h] \centering
\includegraphics[width=0.85\textwidth]{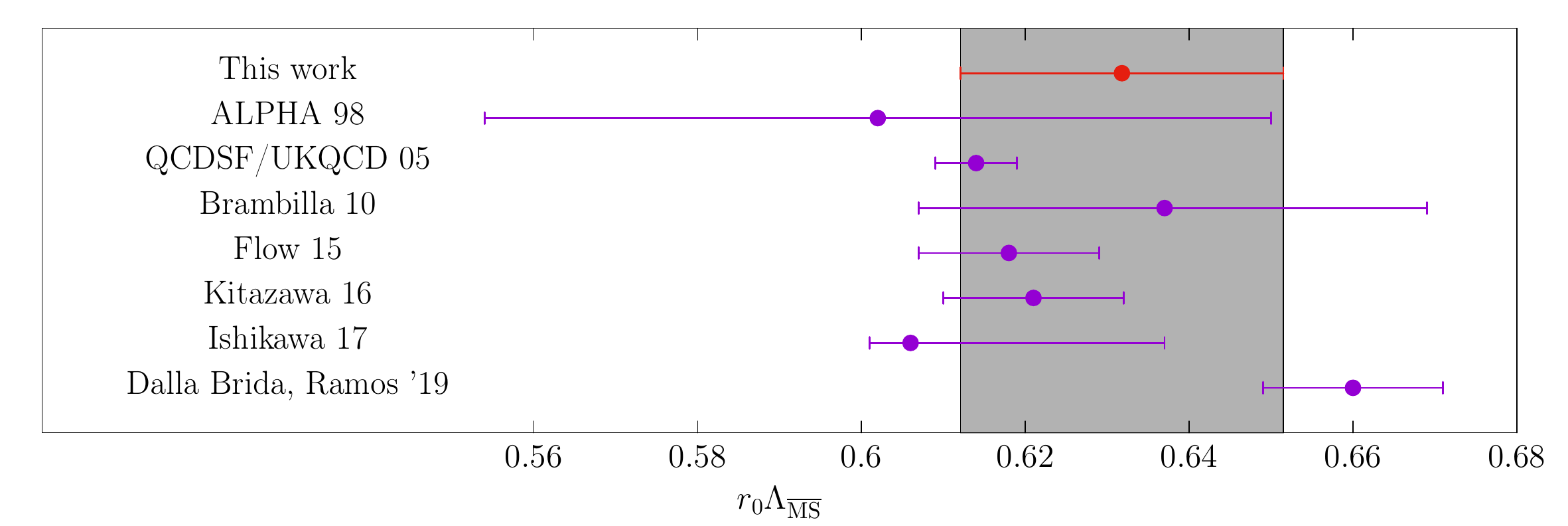}
\caption{Our final result of $r_0\Lambda_{\MS}$ for $SU(3)$~\cite{Bribian:2021cmg} compared with the values included in the FLAG average~\cite{FlavourLatticeAveragingGroup:2019iem}.}
\label{Figure_LMS}
\end{figure}
\vspace{-0.1cm}

 In the case of $SU(3)$, we have used the available data for the Sommer radius $r_0$ as intermediate scale, and fig.~\ref{Figure_LMS} shows the comparison between our result and several determinations for $r_0 \times \Lambda_\MS$ ~\cite{Capitani:1998mq,Gockeler:2005rv,Brambilla:2010pp,Asakawa:2015vta,Kitazawa:2016dsl,Ishikawa:2017xam} reported in the FLAG average~\cite{FlavourLatticeAveragingGroup:2019iem}. Up to this level of precision, our final result $r_0 \times \Lambda_\MS= 0.632(20)$~\cite{Bribian:2021cmg} turns out to be compatible with both the FLAG average  $r_0 \times \Lambda_\MS= 0.615(18)$ and the result by Dalla Brida and Ramos $r_0 \times \Lambda_\MS= 0.660(11)$~\cite{DallaBrida:2019wur}. At this stage, our result is not precise enough to clarify the tension among the existing results in the literature but due to the reduced memory footprint, we believe that this scheme can be efficiently used to pin down the errors.

\section{Dependence on the number of colours}
\label{sec:finite}
\label{finite}
\begin{figure}[t]
\centering
\includegraphics[width=0.8\textwidth]{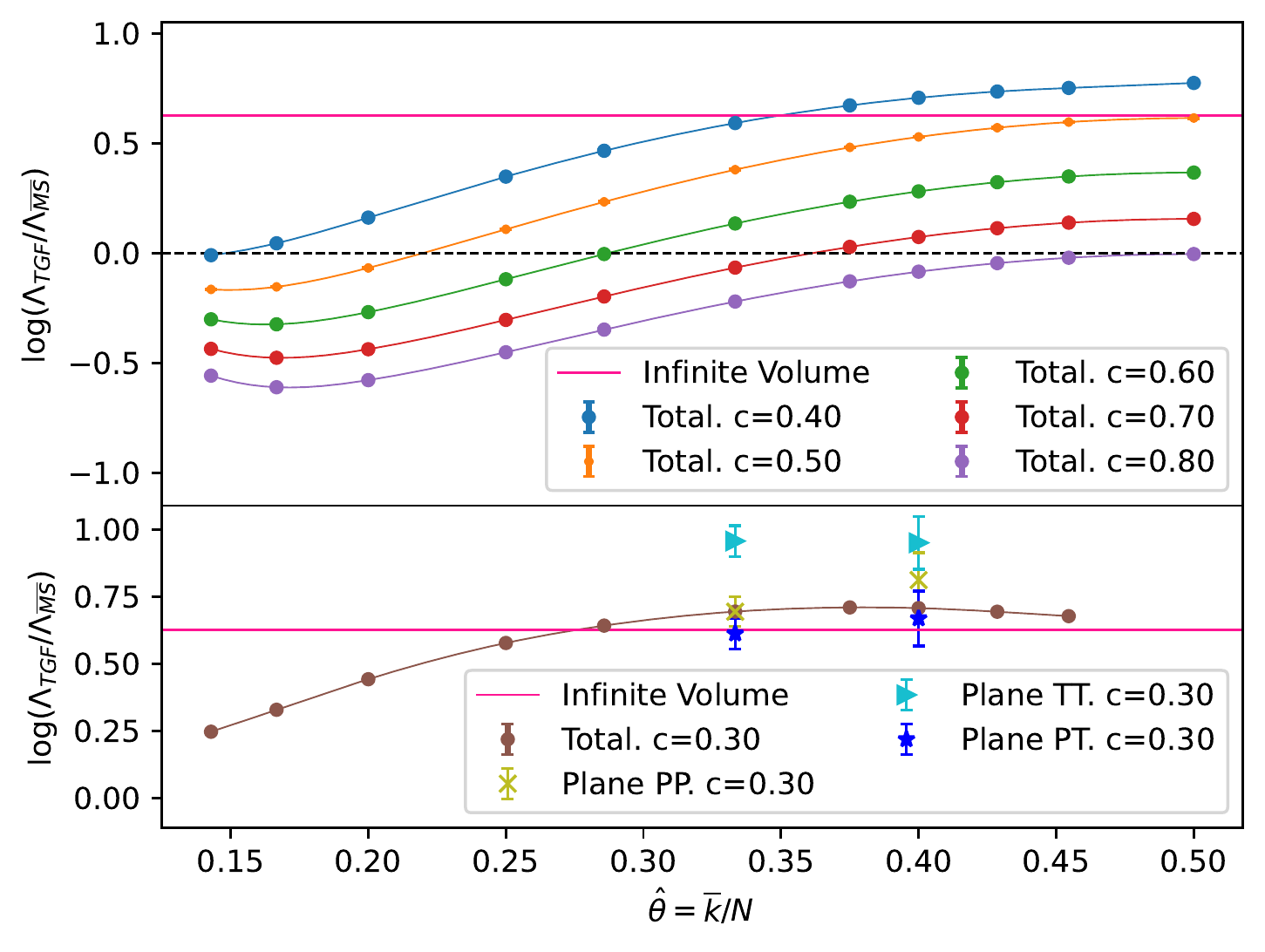}
\caption{We display $\log\left(\LTGF/\Lambda_\MS\right)$ as a function of $\hat{\theta} \equiv \bar{k}/N$ for different choices of $N$ and k-twist. The determination labelled as total is extracted analytically from the perturbative computation of \cite{Bribian:2019ybc}, and the per-plane ratios are computed numerically as explained in the text.}
\label{eduardo}
\end{figure}

In this section, we will describe our ongoing calculation of the $\Lambda-$parameter as a function of the number of colours. Figure~\ref{eduardo} shows the dependence of the ratio $\log\left(\LTGF/\Lambda_\MS\right)$ as a function of $\hat{\theta} \equiv \bar{k}/N$ with $k\bar{k}=1(\text{mod }N$), a parameter that can be directly related to the non-commutativity parameter in non-commutative gauge theories. Most data of this plot is obtained analytically from the perturbative computation of \cite{Bribian:2019ybc}, and we also have included the $c=0.30$ data not available in that reference. For the gauge groups studied in this work, the choice of $N$ and $k$ is not arbitrary, and their values are selected so as to avoid the appearance of tachyonic instabilities in the large $N$ limit~\cite{Perez:2018afi,Chamizo:2016msz}. In order to saturate the bound that enforces the absence of these instabilities, the value of $k$ and the number of colours of the non-abelian group should follow a special and peculiar relation; more precisely, they are chosen respectively as the $n-2$ and $n$-th terms of the Fibonacci sequence, i.e.  $k = F_{n-2}$ and $N = F_n$ , for any value of $n$. In the large $N$ limit, the non-commutativity parameter tends to a fixed value $\hat{\theta}=\varphi^{-2}\simeq0.38196601$, where $\varphi$ is the Golden Ratio. 

\begin{figure}th] \centering
\begin{subfigure}{.5\textwidth} \centering
\includegraphics[width=\linewidth]{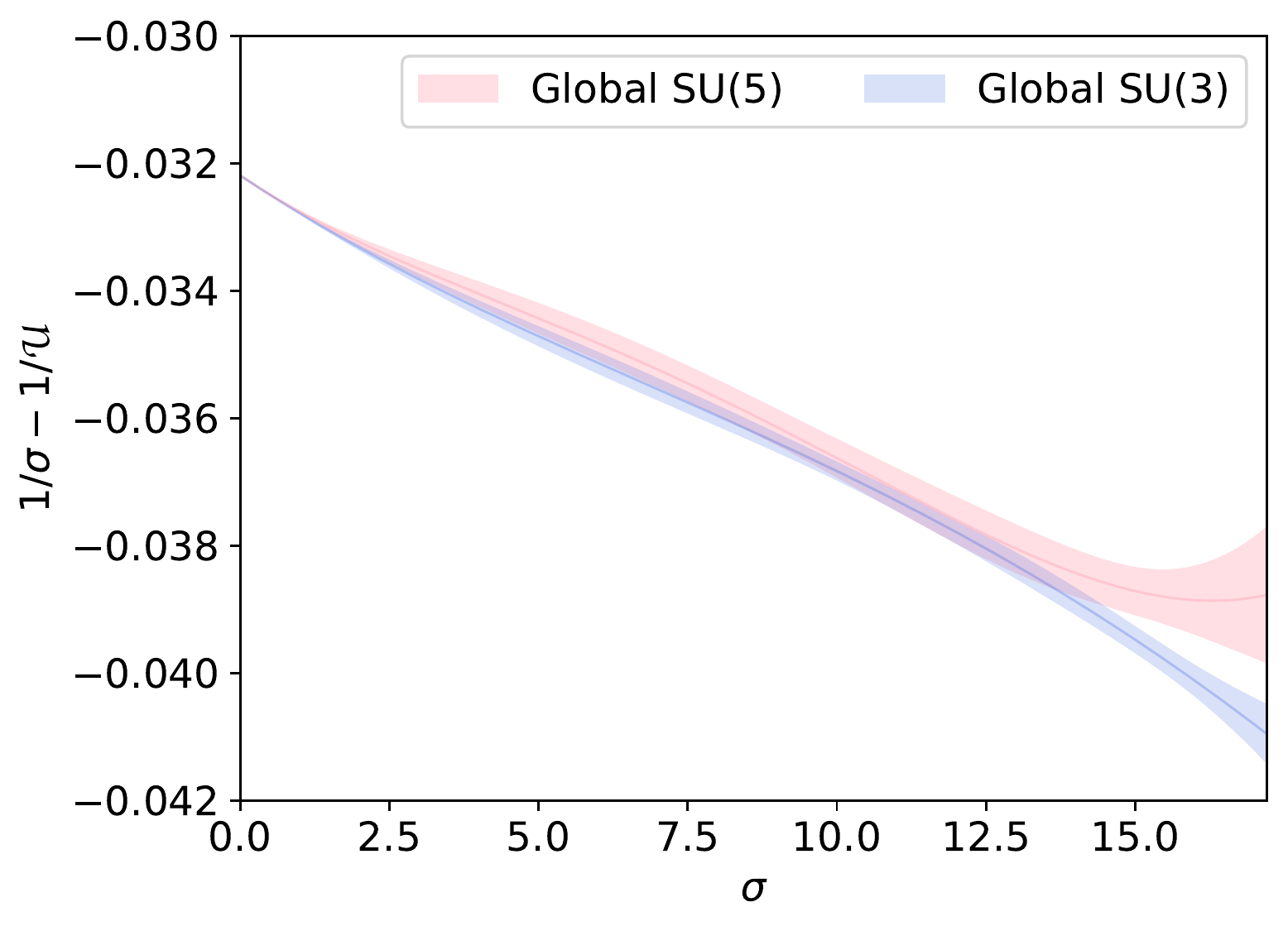}
\caption{Raw data for $1/\sigma - 1/{\cal U}(\sigma, \tL)$ vs
$\sigma$.}
\label{Figure_matching_su303}
\end{subfigure}%
\begin{subfigure}{.5\textwidth} \centering
\includegraphics[width=\linewidth]{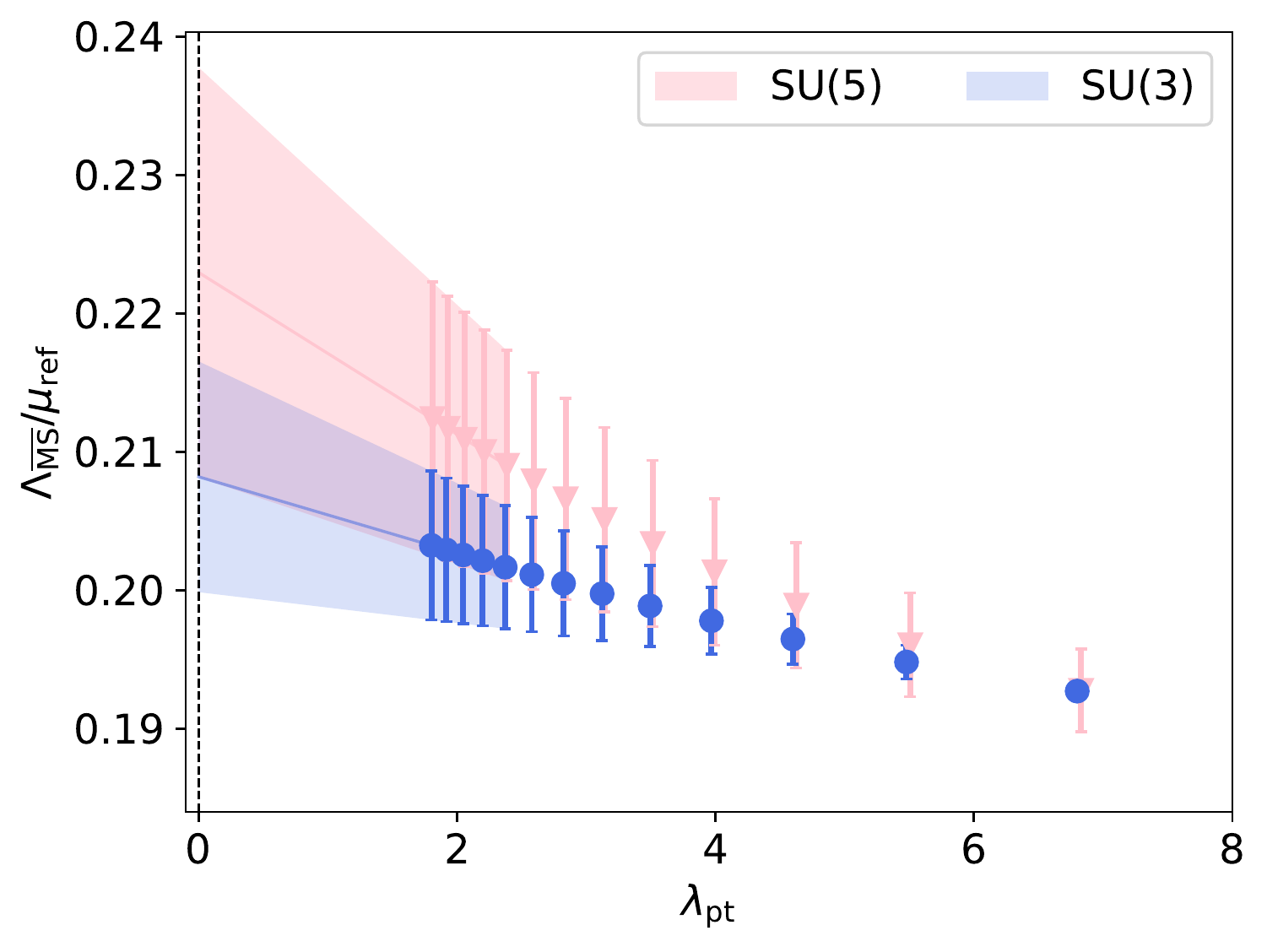}
\caption{Continuum extrapolation of $\Lambda_{\text{TGF}}/\mu_{\text{ref}}$.}
\label{Figure_matching_su3}
\end{subfigure}
\caption{(a) Continuum extrapolation of the inverse step scaling function for $SU(3)$ and $SU(5)$. (b) Continuum extrapolation of $\Lambda_{\text{TGF}}/\mu_{\text{ref}}$ for $SU(5)$ compared to the one of $SU(3)$.}
\label{lambdasu5}
\end{figure}

We start this analysis by computing $\Lambda_{\MS}/\mu_{\text{ref}}$ for the first steps in the Fibonacci sequence, including $(\text{k},N)=(1,3)$, $(2,5)$ and $(3,8)$ (the calculation for $SU(8)$ is ongoing). We will take the large $N$ limit at fixed values of the renormalized 't Hooft coupling $\lambda(\mu_{\text{ref}})\equiv 13.9164955$, thus fixing in all cases the same renormalisation condition. The results for the step scaling function $\sigma(u)$ and $\Lambda_{\MS}/\mu_{\text{ref}}$ for $SU(3)$ and $SU(5)$ are shown in fig.~\ref{lambdasu5}, and collected in table~\ref{table1_planes}. The two determinations come out to be compatible within errors. 

\subsection{Dependence on the boundary conditions}
Although perturbation theory indicates that, in the large $N$ limit, the box is effectively symmetric with size $\tilde{l}$, the boundary conditions in the twisted and untwisted planes are different. To analyse the influence in the determination of the coupling, we can separate the computation of the coupling per planes, by averaging only over a specific set of directions. We identify 3 different types of plane:
\begin{itemize}
        \item \textit{TT} -- The plane having a non-trivial twist. In this case both directions are {\it short}, of extent $l$.
        \item \textit{PT} -- Four planes sharing one direction with the twisted plane, with one {\it short} direction of size $l$, and one {\it long} direction
of size $\tl= N l$.
        \item \textit{PP} -- The plane orthogonal to the twisted one. Here both directions are {\it long} with size $\tl$.
\end{itemize}

Notice that boundary conditions in each plane are different, so each plane represents a different renormalisation scheme. Reproducing the computation of $\Lambda_{\text{TGF}}/\mu_{\text{ref}}$ per plane gives an estimation for the ratios $\LTGF(\text{Plane})/\Lambda_\MS$, relating each TGF-plane scheme with the $\overline{\text{MS}}$ scheme. This ratios have not been computed in perturbation theory so this calculation provides a numerical determination. We have done this through a non-perturbative matching between the coupling computed in each set of planes and the coupling computed averaging over all planes. The advantage of this procedure is that the hadronic scale is kept equal in all cases, as the matching is done at the same bare coupling. Then, the ratios are computed as follows:
\begin{equation}
R\equiv\frac{\LTGF(\text{Plane})}{\Lambda_\MS}=\frac{\Lambda_{\text{TGF}}(\text{Plane})/\mu_{\text{ref}}}{\Lambda_{\text{TGF}}(\text{All})/\mu_{\text{ref}}}\times\frac{\Lambda_{\text{TGF}}(\text{All})}{\Lambda_\MS}\,,
\label{eq:ratio}
\end{equation}
where the last factor refers to equation~\eqref{eq:LTGF_LMS}. 

Figure~\ref{Figure_matching_su3} shows the non-perturbative matching between the planes and the total coupling, taking the PP computation as an explicit example. The purple band is a continuum extrapolation based on the following functional form:
\begin{equation}
  \frac{1}{\lambda_{\text{Plane}}(\mu)} - \frac{1}{u(\mu)} = \sum_{n=0}^{3} c_k u^k +
  \left( \frac{1}{\tilde L^2}\right) \times
  \sum_{n=0}^{8} \rho_k u^k,
\end{equation}
with $u(\mu)=\lambda_{\text{Total}}(\mu)$. Table~\ref{table1_planes} contains the results for $\Lambda_{\text{TGF}}(\text{Plane})/\mu_{\text{ref}}$ and $R=\LTGF(\text{Plane})/\Lambda_\MS$.
\begin{figure}[t] \centering
\begin{subfigure}{.5\textwidth} \centering
\includegraphics[width=\linewidth]{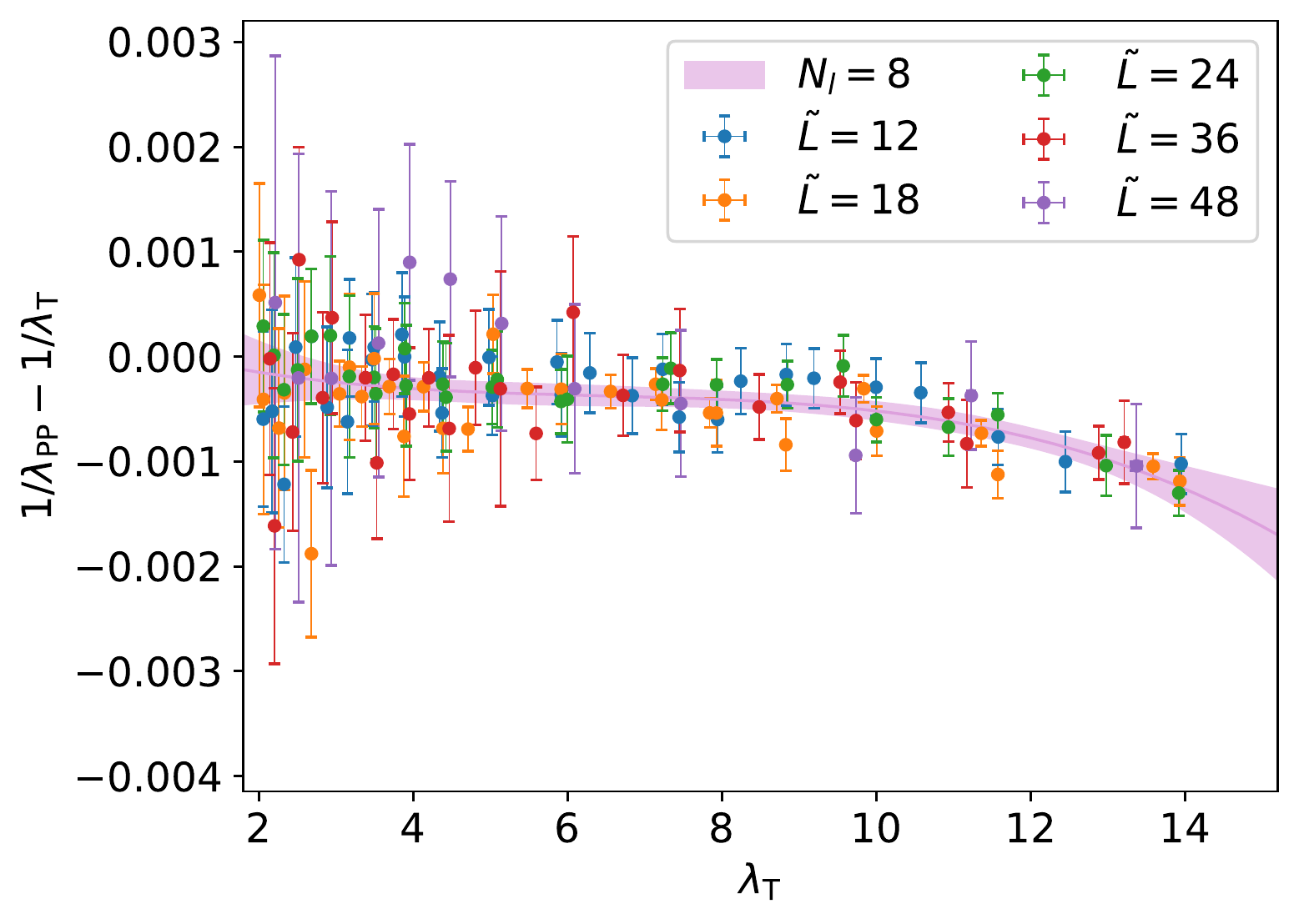}
\caption{SU(3).}
\label{Figure_matching_su303}
\end{subfigure}%
\begin{subfigure}{.5\textwidth} \centering
\includegraphics[width=\linewidth]{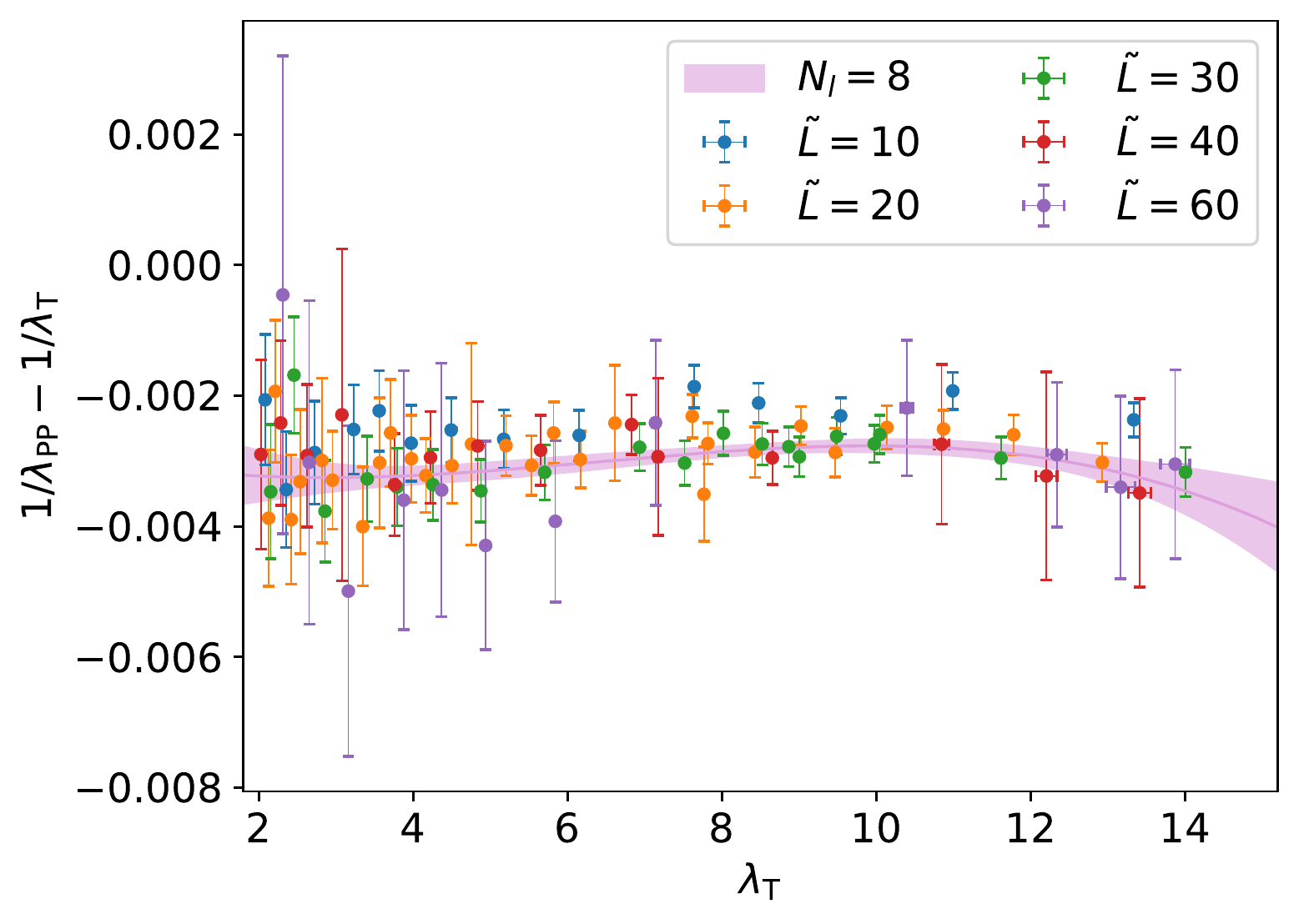}
\caption{SU(5).}
\label{Figure_matching_su3}
\end{subfigure}
\caption{Non-perturbative matching between PP and total coupling $\lambda$ for: (a) $SU(3)$ and (b) SU(5).}
\label{Figure_matching_su3}
\end{figure}

\begin{table}[h]
\centering
\begin{tabular}{ c | c |c | c \Vert c | c | c | c  }
N& Plane & $\Lambda_{\text{TGF}}/\mu_{\text{ref}}$ & R & N&Plane & $\Lambda_{\text{TGF}}/\mu_{\text{ref}}$ & R  \\
 \hline
3 & All& 0.418(17) & 2.0063(11)  & 5 & All& 0.436(33)&  2.0309(39) \\
3 & TT & 0.543(23) & 2.60(15)    & 5 & TT & 0.545(44)&  2.59(25)\\
3 & PT & 0.385(16) & 1.84(11)    & 5 & PT & 0.465(34)&  1.95(20)\\
3 & PP & 0.417(16) & 2.00(11)    & 5 & PP & 0.401(31)&  2.25(23)\\

\end{tabular}
\caption{Results for  $\Lambda_{\text{TGF}}/\mu_{\text{ref}}$ for $SU(3)$ and $SU(5)$ extracted from the average over all planes or over TT, PP or PT planes. We also show the values of the R ratio from equation eq.~(\ref{eq:ratio}).}
\label{table1_planes}
\end{table}

The values of $\log\left(\LTGF(\text{Plane})/\Lambda_\MS\right)$ for the two cases $\hat{\theta}=1/3$ and $\hat{\theta}=2/5$ studied in this work are displayed in fig.~\ref{eduardo} compared with the determinations corresponding to averaging over all planes. For finite volumes, even if twisted boundary conditions successfully replicates the small size of the box, and the torus has an effective volume $\tilde{L}^4$, the boundary conditions in each plane are different and the twisted plane has a non trivial non-commutativity parameter; even so, fig.~\ref{eduardo} shows no large deviations between the computation in each plane and the total one. We are at the present exploring more values of $N$ along the Fibonacci sequence. 
\subsection{Finite volume effects}
\begin{figure}[t] \centering
\begin{subfigure}{.5\textwidth} \centering
\includegraphics[width=\linewidth]{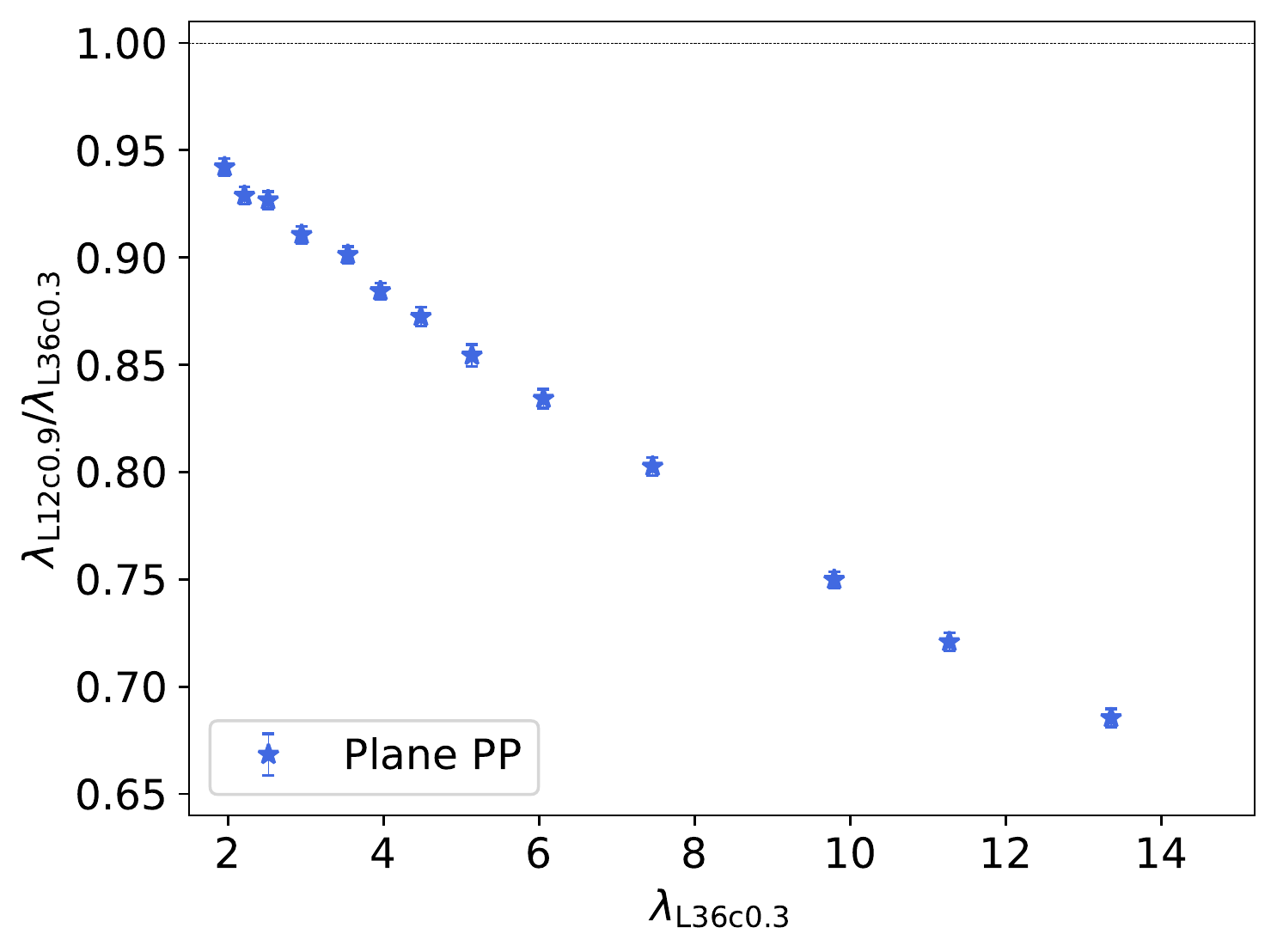}
\caption{$\tilde{L}=12|_{c=0.9}\text{ VS }\tilde{L}=36|_{c=0.3}$.}
\label{0.90y0.30}
\end{subfigure}%
\begin{subfigure}{.5\textwidth} \centering
\includegraphics[width=\linewidth]{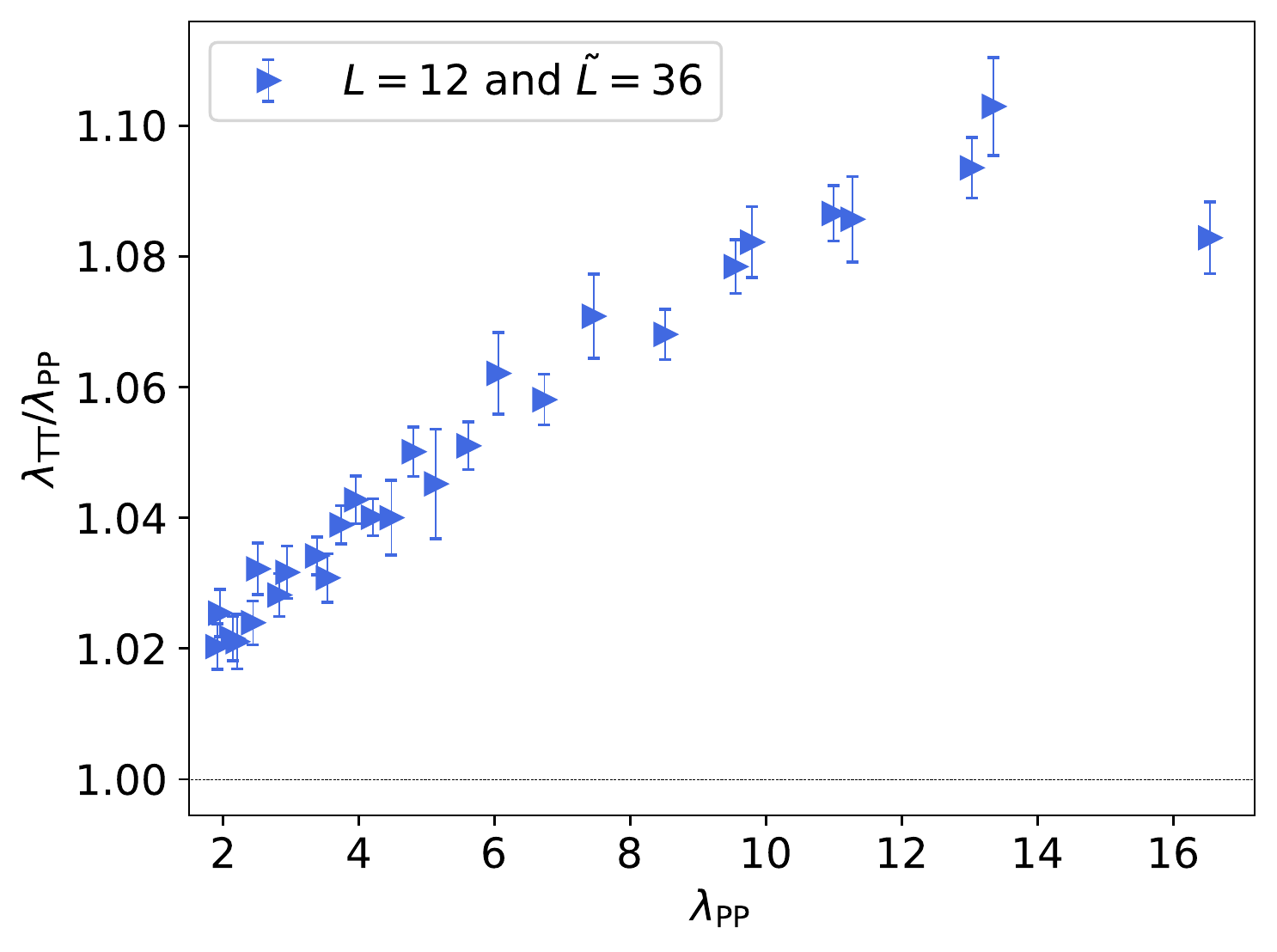}
\caption{$\lambda_{\text{TT}}\text{ VS }\lambda_{\text{PP}}$.}
\label{TTyPP}
\end{subfigure}
\caption{Comparison between finite volume effects in TT and PP computation with. respect to the regular set-up for large values of $c$.}
\label{finite_volume}
\end{figure}

Finally, we will try to evaluate if the twisted lattice succeeds in reducing finite size effects and exhibits an effective volume dependence with $\tilde{l}$. For that purpose, we will compare the computation of the coupling in the TT and PP planes with the standard set-up at large values of $c$. We have compared the renormalized coupling in PP planes obtained at $c=0.30\times N$ for size $L$ with the one obtained for $c=0.3$ and size $N\times L$. In the absence of finite volume effects, these two coupling should be equal. We computed, at fixed value of the bare coupling, the ratio between the $SU(3)$ PP renormalized coupling on a lattice with $\tilde{L}=12$ at $c=0.90$ and $\tilde{L}=36$ ant $c=0.30$, so formally the bare coupling and the renormalization scale are kept constant. Figure~(\ref{0.90y0.30}) displays the ratio of the two couplings, showing that finite volume effects amount to 30\% at strong coupling. On the other hand, fig.~(\ref{TTyPP}) displays the ratio between TT and PP computation for $c=0.30$. In this case, for the TT plane, the renormalisation scale is $\mu^{-1}=c\tilde{l}=\tilde{c}l$, with $\tilde{c}=0.9$, so in principle one could expect a 30\% difference between TT and PP couplings due to finite volume effects. Figure.~(\ref{TTyPP}) shows the deviation at strong coupling is at most 10\%, indicating that twisted boundary conditions indeed succeed in reducing finite volume effects. The study of the $N$ dependence of this effects is ongoing. 
\vspace{-0.27cm}
\vspace{-0.5cm}

\section{Summary}
\label{sec:con}
We have reported on the ongoing investigation of the running coupling of $SU(N)$ YM theories for several values of $N$ using the TGF scheme, that combines three main ingredients: a coupling based on the GF, twisted boundary conditions and an asymmetric geometry. With this scheme, we have determine $\Lambda_{\rm \MS}/\mu_{\text{ref}}$ for $SU(3)$, obtaining good agreement with the literature. We have also presented our ongoing analysis of the $N$ dependence of the $\Lambda-$parameter, for which the TGF scheme is particularly suited. 

\section*{Acknowledgments}
\addcontentsline{toc}{section}{Acknowledgments}

AR acknowledges support from the Generalitat Valenciana
(genT program CIDEGENT/2019/040). JDG and MGP acknowledge support from the MINECO/FEDER grant PGC2018-094857-B-I00 and the MINECO Centro de Excelencia Severo Ochoa
Program SEV-2016-0597. JDG acknowledges support under the FPI grant
PRE2018-084489.  This publication is supported by the project
H2020-MSCAITN-2018-813942 (EuroPLEx) and the EU Horizon 2020 research
and innovation programme, STRONG-2020 project, under grant agreement
No 824093. We acknowledge of the Hydra cluster at IFT and CESGA (Supercomputing Centre of Galicia).

\end{document}